\documentclass[prl,12pt]{revtex4}

\usepackage[english]{babel}
\usepackage[intlimits]{amsmath}
\usepackage[dvips]{graphicx}
\begin{document} 

\title{Exact Solutions of Burnt-Bridge Models for Molecular Motor Transport}
\author{Alexander Yu. Morozov, Ekaterina Pronina,  and Anatoly B. Kolomeisky}
\affiliation{Department of Chemistry, Rice University, Houston, TX 77005 USA}

\author{Maxim N. Artyomov}
\affiliation{Department of Chemistry, Massachusetts Institute of Technology, Cambridge, MA 02139 USA}

\begin{abstract}

Transport of molecular motors,  stimulated by  interactions  with specific links between consecutive binding sites (called ``bridges''), is investigated theoretically by analyzing discrete-state stochastic ``burnt-bridge'' models. When an unbiased diffusing  particle crosses the bridge, the link can be destroyed (``burned'') with a probability $p$, creating a biased directed motion for the particle. It is shown that for  probability of burning  $p=1$ the system can be mapped into  one-dimensional  single-particle hopping model along the  periodic infinite lattice that allows one to calculate exactly all dynamic properties. For general case of $p<1$ a new theoretical method is developed, and  dynamic properties are computed explicitly. Discrete-time and continuous-time dynamics, periodic and random distribution of bridges  and different burning dynamics are analyzed and compared. Theoretical predictions are supported by extensive Monte Carlo computer simulations. Theoretical results are applied for analysis of the experiments on collagenase motor proteins.

\end{abstract}

\maketitle

\section{Introduction}

Molecular motors, or motor proteins, play  important role in the functioning of biological systems \cite{lodish_book,howard_book,bray_book}. They transform a chemical energy into mechanical work with a high efficiency and move along linear molecular tracks. Conventional molecular motors fuel their motion by hydrolyzing adenosine triphosphate (ATP) or related compounds, and the chemical energy influx creates a biased directed motion along one-dimensional protein  filaments.  During the translocation of motor proteins the chemical state of  linear protein tracks remains unchanged. 

Recently, a new type of molecular motor, independent of ATP hydrolysis, has been discovered \cite{saffarian04,saffarian06}. A special protein called collagenase moves along collagen fibrils, which are the main component of extracellular matrix \cite{lodish_book}, and it cleaves the filament at special sites. Collagen proteolysis provides a chemical energy that produces a directed motion of the motor protein molecule. It was found that the random motion of enzyme molecule can be transformed into the biased diffusion along the linear track if, after the cleavage of collagen, the motor protein could only move on one side of the broken link \cite{saffarian06}. 

It was suggested that the dynamics of collagenases can be explained by a so called ``burnt-bridge'' model (BBM) \cite{ saffarian04,saffarian06,mai01,antal05}. According to this approach, the motor protein is viewed as a random walker hopping along the one-dimensional lattice that consists of two types of links: strong and weak. Strong links are not affected  when the particle passes them, however crossing  the weak links might destroy them with a probability $0 < p \le 1$, and the random walker is not allowed to move again over the  ``burnt'' links. Using a  spatial continuum approximation, that neglects the underlying lattice, the burnt-bridge model has been discussed  in the limiting cases of very low $p$ and $p \rightarrow 1$ \cite{mai01}. This approximation is only valid in the limit of very low concentrations of the special links, although in biological systems the density of bridges might be significant \cite{saffarian04,saffarian06}. Antal and Krapivsky  \cite{antal05} have investigated BBM by using a discrete-time random walk approach that allowed to calculate exactly velocities and dispersions for some sets of parameters for several versions of the model. However, discrete-time dynamics is unrealistic for motor proteins since their dynamics is tightly coupled to a set of chemical transitions \cite{kolomeisky00}, and  time intervals between these transitions are distributed exponentially according to Poisson statistics. A discrete-time statistics implicitly assumes a delta-function waiting time distribution between consecutive steps. In addition, the approach presented in Ref. \cite{antal05} is mathematically very involved, and not all relevant dynamic properties have been computed. The goal of this paper is to investigate more realistic continuous-time discrete-state stochastic burnt-bridge models using simple analytical approaches and extensive Monte Carlo computer simulations.

\section{Model}

Let us consider  a single random walker that moves along an infinite one-dimensional lattice as shown in Fig. 1. The particle can hop in both directions with the same rate $v_{0}=1$, i.e., initially there is no bias in the motion of the random walker. We set the size of the lattice spacing to be equal to one. There are two types of links between consecutive binding sites on the lattice. The random walker does not interact with the strong links, while crossing the weak link leads to burning of the bridge with the probability $p$.  We assume that after the burning of the bridge the particle is always on the right side from it, and there are no burnt bridges at the initial time. Thus the particle is never trapped between broken links, and it moves continuously forward.

We distinguish two different possibilities of burning the bridge. In the first option, the bridge is burned only when the particle crosses it from left to right, but nothing happens when the walker moves from right to left. This model is called a forward BBM. In the second option, the particle burns the bridge on any forward and backward motion across the weak link, describing a forward-backward BBM. It is important to note that  both cases are identical for $p=1$, while  for $p<1$  dynamics differs significantly for different burning scenarios. Dynamic properties of the system also depend on the details the particle behavior near the already burned bridge \cite{antal05}. In the standard version of BBM the particle sitting next the broken link can only move forward to the right direction. In the modified version of BBM the walker tries to cross the burnt bridge, but the attempt fails and the particle does not change its position. 

The dynamics of the motor protein molecule in  different BBMs  is directly determined by two parameters: the probability of burning $p$ and the concentration of weak links $c$. The distribution of bridges strongly influences the dynamic properties \cite{antal05}, and we will consider two different cases: periodic and random distributions. The case when the bridges are found at equal distance $N=1/c$ lattice spacings apart from each other describes the periodic distribution, while in the random case we postulate that each link is a bridge with the probability $c$ every time the walker visits it. Note, however, that other realizations of the random distribution of bridges are also possible.

\section{Burnt-Bridge Models with $p=1$}

First, we consider a special case of BBM with $p=1$ when any crossing of the bridge leads to its burning. In addition, it is assumed that the first burnt bridge was crossed from the left. Thus the particle drifts to the right (forward) direction. For the periodic distribution of bridges the dynamic properties of the system can be evaluated explicitly by mapping the model into the motion of the random walker on infinite one-dimensional periodic lattice, for which exact solutions have been derived by Derrida in 1983 \cite{kolomeisky00,derrida83,AR07}. The particle at site $j$ ($j=0,1,\cdots,N-1$) moves forward (backward) with the rate $u_{j}$ ($w_{j}$), and the periodicity requires that $u_j\equiv u_{j+N}$ and $w_j\equiv w_{j+N}$, where $N$ is the size of the period. For this model  the explicit expression for the mean velocity $V$ is given by  \cite{derrida83} 
\begin{equation}\label{vel}
V =\frac{N}{\sum\limits_{j=0}^{N-1}r_j} \left[1-\prod\limits_{i=1}^{N}\left( \frac{w_i}{u_i}\right)\right],
\end{equation}
while the dispersion, or diffusion coefficient, can be found from
\begin{equation}\label{D}
D = \frac{1}{\left(\sum\limits_{j=0}^{N-1}r_j\right)^2} \left\{ V \sum\limits_{j=0}^{N-1} s_j\sum\limits_{i=0}^{N-1} (i+1)r_{j+i+1} +N \sum\limits_{j=0}^{N-1} u_j s_j r_j \right\} -\frac{V (N+2)}{2},
\end{equation}
with auxiliary functions
\begin{equation}\label{r_j}
r_j = \frac{1}{u_j}\left[ 1+\sum\limits_{k=1}^{N-1} \prod\limits_{i=j+1}^{j+k}\left(\frac{w_i}{u_i}\right)\right],
\end{equation}
and
\begin{equation}\label{s_j}
s_j = \frac{1}{u_j}\left[ 1+\sum\limits_{k=1}^{N-1} \prod\limits_{i=j-1}^{j-k}\left(\frac{w_{i+1}}{u_i}\right)\right].
\end{equation}

The BBM with $p=1$ corresponds to the particle hopping model with $u_{j}=w_{j}=1$  for all $j$ with the exception of $w_{0}=0$, that reflects the complete burning of the bridge. Then from Eq. (\ref{vel}) the explicit formula for the velocity  is given by  
\begin{equation}\label{vv}
V = \frac{2}{N+1} = \frac{2 c}{c+1}.
\end{equation}
The dispersion  can be obtained from Eq. (\ref{D}), yielding
\begin{equation}
D = \frac{2 (c^2+c+1)}{3 (c+1)^2}.
\end{equation}
The dynamic properties of BBM with $p=1$ are shown in Fig. 2 as  functions of the concentration of bridges. The velocity is a monotonically growing function of $c$, while the dispersion is always decreasing. It is interesting to consider their behavior in limiting cases. When every link on the lattice is a bridge ($c=1$) we obtain $V=1$ and $D=1/2$, as expected. More interesting is the limit of  $c \rightarrow 0$. The velocity  goes to zero, $V \simeq 2c$, as expected, and this reproduces the result of the continuum theory of Mai et al. \cite{mai01}. But, surprisingly, it is found that $D_{BBM}(c=0)=2/3$ and it is  not equal to the dispersion of a free particle on the lattice without bridges, $D_{free}=1$. This jump in the dispersion might correspond to a dynamic transition, indicating qualitatively different behavior of the particle moving in the system without weak links and particle diffusing in BBM. Similar behavior has been observed in the discrete-time BBM \cite{antal05}. However, this interesting phenomenon requires a more careful and detailed investigation. 

The mapping of BBM to  the particle hopping model on the periodic lattice  can also be used to investigate discrete-time burnt-bridge models. It was found \cite{derrida83} that in this case the explicit expression for the velocity is the same as in the continuous case, while for the diffusion coefficient it can be shown that $D_{discr}=D_{cont}-V^{2}/2$, and  $D_{cont}$ and $V$  can be obtained from  Eqs. (\ref{vel}) and (\ref{D}). However, the rate constants $u_{j}$ and $w_{j}$ have to be reinterpreted as jumping probabilities, leading to 
\begin{equation}\label{uw}
u_j = w_j = \frac{1}{2} \quad \mbox { for }   j = 1,2,\cdots,N-1.
\end{equation}
At the same time, for the particle near the burned bridge the rates are $u_{0}=1$ and $w_{0}=0$ for the standard BBM, while $u_{0}=1/2$ and $w_{0}=0$ for the modified BBM. Then the dynamic properties of the standard discrete-time BBM are
\begin{equation}
V=c, \quad D=\frac{1}{3} (1-c^{2}),
\end{equation}
and for the modified BBM we obtain
\begin{equation}
V=\frac{c}{c+1}, \quad D = \frac{1}{3 (c+1)}-\frac{c^2}{6(c+1)^2}.
\end{equation}
These results  fully agree with those obtained in Ref. \cite{antal05} for periodic bridge distribution. Note, however, that our approach based on Derrida's analysis is much more straightforward.

\section{ Burnt-Bridge Models with $p<1$}

Derrida's method cannot be easily extended to include  the general case of bridge burning with the probability $p<1$  since it is not clear how the hopping rates $u_{j}$ and $w_{j}$ depend on $p$. To calculate dynamic properties of BBM we introduce a new method that is very general and applicable for many different systems. Below we analyze in detail several BBM to illustrate our approach.

\subsection{Continuous-time forward BBM for the periodic distribution of bridges}

Now consider a continuous-time BBM with the particle destroying the link with the probability $p$ only when the bridge is crossed from left to right. The distribution of weak links is periodic, i.e., they are positioned $N=1/c$ sites apart from each other. Let us define $R_{j}(t)$ as the probability to find the random walker $j$ sites apart from the last burnt bridge at time $t$. The probabilities $R_{j}(t)$ are associated with the moving system of coordinates  with the last burned bridge always at the origin. The corresponding kinetic scheme is shown in Fig. 3a. The dynamics of the system can be described by a set of Master equations. For the lattice sites not connected to weak links  or connected only from the left side we have
\begin{equation}\label{me1}
\frac{d R_{kN+i}(t)}{dt} = R_{kN+i-1}(t)+R_{kN+i+1}(t)-2 R_{kN+i}(t),
\end{equation}
for $k=0,1,2,\cdots$ and $i=1,2,\cdots,N-1$. For all other lattice sites (except the origin) one can show that
\begin{equation}\label{me2}
\frac{d R_{kN}(t)}{dt}=(1-p)R_{kN-1}(t)+R_{kN+1}(t)-2R_{kN}(t) \quad \mbox{ for } k=1,2,\cdots.
\end{equation}
The dynamics is different at the origin (see Figs. 1 and 3a) where
\begin{equation}\label{me3}
\frac{d R_{0}(t)}{d t}=p \sum_{k=1}^{\infty} [R_{kN-1}(t)] +R_{1}(t)-R_{0}(t).
\end{equation}
In addition, there is a normalization condition, 
\begin{equation}\label{norm}
\sum_{k=0}^{\infty} \sum_{i=0}^{N-1} R_{kN+i}(t)=1, 
\end{equation}
which is valid at all times.

In the stationary-state limit ($t\rightarrow \infty$) we have $\frac{d}{dt}R_j(t)=0$, and Eqs. (\ref{me1}), (\ref{me2}) and (\ref{me3}) simplify significantly, 
\begin{equation}\label{Ri}
R_{i}=R_{0}- i p S \quad S = \sum_{k=1}^{\infty} R_{kN-1}, \quad \mbox{ for } i=0,1,\cdots,N;
\end{equation}
and
\begin{equation}\label{Rk}
R_{kN+i}=(i+1)R_{kN}-i(1-p)R_{kN-1}  \quad  \mbox{ for }  i=0,1,\cdots,N; \  k=1,2,\cdots.
\end{equation}
To find the solutions of Eqs. (\ref{Ri}) and (\ref{Rk}) we assume that it has the following form 
\begin{equation}\label{beta}
R_{kN+i}=R_0 e^{ak} - iB(k),
\end{equation}
where $a$ and $B(k)$ are unknown parameters. This expression assumes a linear dependence between the weak links and an exponential decrease for consecutive periods. To eliminate $B(k)$, we observe that  Eqs. (\ref{Ri}) and (\ref{Rk})  simultaneously  hold for $i=N$, giving us
\begin{equation}
R_{(k+1)N}=R_0 e^{a(k+1)} =R_{kN+N}= R_0 e^{ak} - N B(k),
\end{equation}
that leads to
\begin{equation}\label{aa}
B(k) = \frac{R_0}{N}e^{ak}(1-e^a).
\end{equation}
To obtain the expression for $a$, we substitute $R_{kN+i}$ from Eq. (\ref{beta}) in the recurrent formula (\ref{Rk}),
\begin{equation}\label{gamma}
R_0 e^{ak} - iB(k) = (i+1)R_0 e^{ak} - i(1-p)\left[R_0 e^{a(k-1)}-(N-1)B(k-1)\right].
\end{equation}
Then using the expression (\ref{aa}) for $B(k)$ and defining $x \equiv e^{a}$, Eq. (\ref{gamma}) eventually reduces to 
\begin{equation}\label{eq_x}
x^2-[2+p(N-1)]x+(1-p)=0.
\end{equation}
Since $R_{kN+i}$ is a decreasing function,  the parameter $a$ is always less than one, and we choose a physically reasonable solution of Eq. (\ref{eq_x}),
\begin{equation}\label{gamma1}
x = 1+\frac{1}{2}p(N-1)-\frac{1}{2}\sqrt{p^2(N-1)^2 + 4pN}.
\end{equation}
Note that generally $0 \le x \le 1$ and $x=0$ for $p=1$, while $x=1$ for $p=0$. The probability to find the particle next to the burnt bridge, $R_{0}$, can be found from the normalization condition (\ref{norm}) and Eq. (\ref{gamma1}), yielding
\begin{equation}\label{eq_R0}
R_0 = \frac{2(1-x)}{N+1+x(N-1)}=\frac{2p}{\sqrt{p^2(N-1)^2 + 4pN}} = \frac{2pc}{\sqrt{p^2(1-c)^2 + 4pc}},
\end{equation}
and the solution for $R_{kN+i}$ in terms of $x$ and $R_{0}$ is finally given as
\begin{equation}\label{gamma2}
R_{kN+i} = R_{0} x^k\left[1-\frac{i}{N}(1-x)\right].
\end{equation}

Exact solutions for the probabilities $R_{j}$ to find the particle at given distance from the last burnt bridge allow us to calculate many dynamic properties of the system. Specifically, the mean velocity of the particle is given by 
\begin{equation}
V = \sum\limits_{j=0}^{\infty}(u_j-w_j)R_j.
\end{equation}
However, since the rates to go forward and backward are the same at all sites except the origin, all contributions to the velocity  cancel out and only $j=0$ term survives,  
\begin{equation}\label{lambda}
V(c,p) = R_{0}=  \frac{2pc}{\sqrt{p^2(1-c)^2 + 4pc}}
\end{equation}
For $p=1$ this equation reduces to $V(c,p=1)=2c/(c+1)$, which agrees exactly with the result (\ref{vv}) derived using the mapping to Derrida's random hopping model. In the limit of $p\rightarrow0$ and $c\rightarrow0$ ($p\ll c$)  Eq. (\ref{lambda}) becomes $V \approx \sqrt{pc}$ as compared to $V \approx \sqrt{2pc}$ in \cite{mai01}. The discrepancy is due to the fact that in our model only forward burning is considered, while in Ref. (\cite{mai01}) the bridge can be burned on both forward and backward motions of the motor protein molecule. The expression for the velocity  (\ref{lambda}) also simplifies in the case of $N=c=1$, giving $V=\sqrt{p}$.

Another useful dynamic property that can be obtained from our analysis is average and average-squared distances between the particle and the last burnt bridge, namely,
\begin{equation}\label{eq_j}
\langle j \rangle=\sum\limits_{j=0}^{\infty}jR_j, \mbox{  and  } \langle j^{2} \rangle=\sum\limits_{j=0}^{\infty}j^{2} R_j.
\end{equation} 
Substituting Eq. (\ref{gamma2}) into the expressions (\ref{eq_j}) we obtain
\begin{equation}\label{nc}
\langle j \rangle=R_0\left\{\frac{N^2 x}{(1-x)^2}+\frac{N^2-1}{6}\right\},
\end{equation}
and
\begin{equation}\label{nd}
\langle j^2 \rangle = R_0\left\{\frac{N^3x(x+1)}{(1-x)^3}+\frac{1}{12}N(N^2-1)\frac{x+1}{1-x}\right\},
\end{equation}
where $x$ and $R_0$ are given by Eqs. (\ref{gamma1}) and (\ref{eq_R0}). For $p=1$ one can show that
\begin{equation}
\langle j \rangle =(N-1)/3, \quad \langle j^{2} \rangle=N(N-1)/6.
\end{equation}
In the limit of $p \ll c$ we have  
\begin{equation}
\langle j \rangle \simeq \sqrt{N/p}, \quad \langle j^{2} \rangle \simeq 2N/p,
\end{equation}
while in the opposite limit of $c \ll p$ the results are
\begin{equation}
\langle j \rangle \simeq N/3, \quad \langle j^{2} \rangle \simeq N^{2}/6.
\end{equation}
For the special case of $N=c=1$ we obtain
\begin{equation}
\langle j \rangle = \frac{1}{\sqrt{p}}-1, \quad \langle j^{2} \rangle = (\frac{1}{\sqrt{p}}-1)(\frac{2}{\sqrt{p}}-1).
\end{equation}

We can also calculate the average and average-squared distances between the consecutive burnt bridges. Because the burning is a stochastic process, i.e., $p <1$, this distance generally deviates from $N$. In order to compute these properties we introduce $P_{kN}$ as a probability that the next bridge that burned is $kN$ sites from the given burnt bridge. Then
\begin{equation}\label{eq_P}
P_{kN}=\frac{p R_{kN-1}}{\sum_{k=1}^{\infty} (p R_{kN-1})}, \quad k=1,2,\cdots.
\end{equation}
The average and average-squared distances between the consecutive burnt bridges can be found from the following expressions
\begin{equation}\label{eq_l}
\langle l \rangle = \sum_{k=1}^{\infty} kN P_{kN}, \quad \langle l^{2} \rangle = \sum_{k=1}^{\infty} (kN)^{2} P_{kN}.
\end{equation}
By substituting Eqs. (\ref{gamma2}) and (\ref{eq_P}) into (\ref{eq_l}) and using Eq. (\ref{gamma1}) we found that
\begin{equation}
\langle l \rangle = \frac{N}{1-x}, \quad \langle l^{2} \rangle =\frac{N^{2}(1+x)}{(1-x)^{2}}.
\end{equation}
The average distance $\langle l \rangle $ is related to the fraction of unburnt bridges $q$ via the following expression
\begin{equation}
q=\frac{(\langle l \rangle/N - 1)}{\langle l \rangle/N}= 1- N/\langle l \rangle=x,
\end{equation}
which gives the physical meaning of the variable $x$ used in our analysis.
In the limiting cases we obtain the following results:
\begin{equation}
\begin{array}{lll}
\langle l \rangle=N, & \langle l^{2}\rangle = N^{2}, & \mbox{ for } p=1; \\
\langle l \rangle=1/\sqrt{p}, &  \langle l^{2} \rangle = 2/p-1/\sqrt{p}, & \mbox{ for } N=c=1; \\
\langle l \rangle= \sqrt{N/p}, &  \langle l^{2} \rangle = 2N/p,  & \mbox{ for } p \ll c;  \\
\langle l \rangle= N, & \langle l^{2}\rangle = N^{2}, & \mbox{ for } p \gg c.
\end{array}
\end{equation}

\subsection{Discrete-time forward-backward BBM for the periodic distribution of bridges}

Our theoretical method  is very general and it can be applied to any version of BBM. As another example, we analyze a motion of the particle in discrete-time framework with periodically located bridges that  can burn with the probability $p$ during the forward and backward transitions. We again introduce a modified kinetic scheme by putting the last burnt bridge at the origin, as shown in Fig 3b. The important difference between continuous-time and discrete-time models is the fact that instead of transition rates in the continuous-time case transition probabilities are used in the discrete-time dynamics. As a result, the sum of transition probabilities at each site is equal to unity. We assume that at the sites that do not support the weak links the transition probabilities are equal, i.e., $u_{kN+i}=w_{kN+i}=1/2$ and $w_{kN-1}=u_{kN}=1/2$ for any $k$ and for $i=1,2,\cdots,N-2$. The transition probabilities are also equal at the sites that surround the weak links, giving $u_{kN-1}=w_{kN}=(1-p)/2$ for $k=1,2,\cdots$ (see Fig. 3b). In addition, at the origin it is assumed  that $u_{0}=1$ and $w_{0}=0$.

As before, we define $R_{j}(t)$ as the probability to find the random walker at the position $j$ at time $t$. The dynamics of the system is governed by a set of discrete-time Master equations that are different from the continuous-time version \cite{antal05,derrida83}. For the lattice sites that do not support the weak links the corresponding Master equations are
\begin{equation}\label{eqn1}
R_{kN+i}(t+1)=\frac{1}{2} R_{kN+i-1}(t)+\frac{1}{2} R_{kN+i+1}(t),
\end{equation}
for any $k$ and with $i=1,2,\cdots,N-2$ (with the exception of $R_{0}$ and $R_{1}$ --- see below). For the lattice sites around the bridges we have
\begin {eqnarray}\label{eqn2}
R_{kN-1}(t+1) & = & \frac{1}{2} R_{kN-2}(t)+\frac{1-p}{2} R_{kN}(t),   \nonumber \\
R_{kN}(t+1) & = & \frac{1-p}{2} R_{kN-1}(t)+\frac{1}{2} R_{kN+1}(t),
\end{eqnarray}
for $k=1,2,\cdots$. The dynamics is different near the origin,
\begin {eqnarray}\label{eqn3}
R_{0}(t+1) & = & \frac{p}{2} S^{\prime} + \frac{1}{2} R_{1}(t),   \nonumber \\
R_{1}(t+1) & = &  R_{0}(t)+\frac{1}{2} R_{2}(t),
\end{eqnarray}
where $S'= \sum_{k=1}^{\infty} [R_{kN-1}(t)+R_{kN}(t)]$. At large times the system reaches a stationary state with $R_{kN+i}(t+1)=R_{kN+i}(t)$, and Eqs. (\ref{eqn1}), (\ref{eqn2}) and (\ref{eqn3}) simplify to yield for the first period ($k=0$),
\begin {eqnarray}\label{eqn4}
R_{i}&=&2 R_{0}- ip S^{\prime} \quad \mbox { for } i=1,2,\cdots,N-1; \nonumber \\
R_N&=&\frac{1}{1-p}(2R_0-N p S^{\prime}),
\end{eqnarray}
while for all other periods ($k \ge 1$) one can obtain
\begin {eqnarray} \label{eqn5}
R_{kN+i} &=&(i+1)R_{kN}-i(1-p)R_{kN-1},  \quad \mbox{ for }  i=0,1,...,N-1; \nonumber \\
R_{kN+N} &=&\frac{1}{1-p}\left[(N+1)R_{kN}-N(1-p)R_{kN-1}\right].
\end{eqnarray}

In order to solve Eqs. (\ref{eqn4}) and (\ref{eqn5}) it is natural to look for a solution in the form
\begin{equation}\label{eqn_sol}
R_{kN+i}=2 R_{0} e^{ak} -iB(k),
\end{equation}
for all $k$ and $i$ except the origin ($k=i=0$). Note that the factor of $2$ in front of $R_0$  in Eq. (\ref{eqn_sol}) is necessary to ensure the consistency between the solutions for $k=0$ from Eq. (\ref{eqn4}) and for $k \ge 1$ [see Eq. (\ref{eqn5})]. The solutions can be obtained using the same approach as was discussed in the previous section for the continuous-time  forward BBM. The final expressions are
\begin{eqnarray}\label{eq_R}
R_0&=&\frac{1-x}{N+[N-p(N-1)]x}, \nonumber \\
R_{kN+i}&=&2R_0 x^k \left[1-\frac{i}{N}(1-(1-p)x)\right],
\end{eqnarray}
where we again define $x \equiv e^{a}$, and it can be shown that
\begin{equation}\label{eqn_x}
x=\frac{1}{2}\left(\alpha -\sqrt{\alpha^2-4}\right), \quad \alpha=\frac{N+1}{1-p}-(1-p)(N-1).
\end{equation}

The dynamic properties of the particle in this model can be calculated as presented above for the continuous-time forward BBM. For the velocity we obtain the following formal expression, \begin{equation}\label{eqn_vel}
v=R_0+\frac{p}{2}\sum\limits_{k=1}^{\infty}R_{kN}.
\end{equation}
The validity of this equation can be seen from the fact that the forward and backward hopping rates cancel each other on all sites except at the origin and the sites labeled $kN$ ($k=1,2,\cdots$). From Eq. (\ref{eq_R}) it follows that $R_{kN}=2R_0 x^k$. Substituting this result into (\ref{eqn_vel})  yields 
\begin{equation}
v=R_0\frac{1+(p-1)x}{1-x},
\end{equation}
that can be transformed with the help of Eq. (\ref{eq_R}) into
\begin{equation}
v=\frac{1-(1-p)x}{N+[N-p(N-1)]x}.
\end{equation}
Finally, replacing $x$ according to (\ref{eqn_x}) and using $N=1/c$ we obtain
\begin{equation}\label{lw}
v(c,p)=\frac{c(1-p)(2c+Z)}{2c(1-p)-[1-p(1-c)]Z},
\end{equation}
where
\begin{equation}\label{lw1}
Z=[2c+p(2-p)(1-c)]\left\{-1+\sqrt{1-\frac{4c^2(1-p)^2}{[2c+p(2-p)(1-c)]^2}}\right\}.
\end{equation}
After some rearrangement, it can be shown that this formula is identical to the expression for the velocity obtained by Antal and Krapivsky \cite{antal05}. But note again that our derivation is much simpler.

\subsection{Random distribution of bridges.}

So far we considered  BBM with weak links positioned at equal distances from each other, i.e., periodic distribution of bridges. Our approach also allows  to easily calculate  the properties of the systems with random distribution of bridges. In this case, each link can be a bridge with the probability  $c$ every time the particle visits it, and the link can also burn with the probability $p$. This suggests that BBM with random distribution of bridges is equivalent to BBM with periodic distribution for $N=1$ and with the probability of burning equal to $pc$. Then all dynamic properties can be calculated as explained above by modifying already derived results with $c \rightarrow 1$ and $p \rightarrow pc$. For example, the velocity of the particle in the continuous-time forward BBM with random distribution of bridges is obtained from Eq. (\ref{lambda}) as
\begin{equation}
V_{rand}(c,p)=V_{per}(1,cp)=\sqrt{pc}.
\end{equation}
Also for the random distribution of bridges in the continuous-time forward-backward model the calculations based on the method outlined above produce the following expression,
\begin{equation}
V_{rand}(c,p)=\sqrt{pc(2-pc)}.
\end{equation}
In the limit of $c \ll 1$ and $p \ll 1$ it reduces to $\sqrt{2pc}$, which is in exact agreement with the result obtained for this case in the Ref. \cite{mai01}. It should be noted again,  that there are different possibilities of introducing randomness in the distribution of bridges that differ from our presentation, and that could effect the dynamic properties of the system. However, our theoretical approach can be applied for these cases too.

\section{Discussions}

To check the validity of our theoretical approach we performed extensive Monte Carlo computer simulations for several BBMs. A general algorithm for continuous-time and discrete-time random walks, discussed in detail in Ref. \cite{mccarthy93}, has been utilized. Velocities have been calculated from the single simulations, consisting of $10^{6}$-$10^{7}$ steps, via the basic formula $V=\langle x \rangle/t$. Diffusion constants have been determined using the expression $D=(\langle x^{2} \rangle -\langle x \rangle^{2})/2t$, and typically more than $10^{5}$ simulation runs have been averaged over to reduce the stochastic noise in the data. The results of simulations, as shown in Figs. 2, 4, 5A and 6, are in all cases in excellent agreement with  available analytical calculations, supporting our analytical predictions.

Theoretical calculations show that the velocity of the particle in BBM is a monotonously increasing function of the concentration of weak links  and burning probability: see Figs. 2, 4 and 5. However, the behavior of the dispersion is more complex, as can be judged from Figs. 2 and 6.  For  fixed values of $p$ it  decreases as a function of $c$, although note large fluctuations when $c \rightarrow 0$. Increase in the burning probability first lowers significantly the dispersion, and then it starts to increase slowly. Our computer simulations indicate (see Fig. 6a) that when every link in the system is a potential bridge ($c=1$) the dispersion is always equal to 1/2 for any $p \ne 0$.

Dynamic properties of the particle in BBM depend on the distribution of bridges, as illustrated in Fig. 5A. Contrary to naive expectations, at the same concentration $c$ and the burning probability $p$,  the random walker moves faster for the random and not for the periodic distribution of bridges, although the difference decreases for lower values of $p$. This surprising observation can be explained using the following arguments. When burning of the bridge can take place with a significant probability, the rate-limiting step in the forward motion of the particle is to find the weak link. For the case of the random distribution, when any link can turn out to be a bridge  with the probability  $c$, the time to find the bridge is of the order of $1/c$. In the periodic distribution of bridges the particle has to diffuse a distance $N=1/c$ before reaching the bridge, and the diffusion time is of the order of $N^{2}=1/c^{2}$. Thus, the particle moves faster in the random distribution of bridges. 

Our theoretical calculations also show that details of burning process (forward versus forward-backward) are affecting the dynamics of the particle. Under the same conditions, the random walker moves faster in the forward-backward BBM, and this effect is significant for large $p$ and $c$: see Fig. 5B. These results are expected since the burning of the bridge fuels the forward motion of the particle. The possibility of burning in the forward and backward motion  accelerates the motion of the random walker.

Developed theoretical method allows us to analyze experimental data on the motion of collagenases motor proteins \cite{saffarian04}. Using fluorescence correlation spectroscopy it was determined that the diffusion coefficient is $D= 8 \pm 1.5 \times 10^{-13}$ m$^{2}$ s$^{-1}$  and the mean drift velocity is $V=4.5 \pm 0.36 \times 10^{-6}$ m  s$^{-1}$ \cite{saffarian04}. Our theoretical results are obtained assuming that the lattice spacing and transition rates are equal to unity. To restore the correct units we assume that the motor protein moves in discrete steps of size $a$. Although the exact value of $a$ is unknown we can reasonably take $a \simeq 10$ nm, which is of the same order as step sizes of other molecular motors \cite{howard_book}. From the structure of the collagen it is known that cleavage sites, that correspond to the bridges, are separated by the distance $\Delta =300$ nm \cite{lodish_book,saffarian04}. It suggests that BBM with periodic distribution of bridges and $c=a/\Delta=1/30$ is appropriate to describe the dynamics of the system. Since at low concentrations the predictions for the dynamic properties in the forward and forward-backward BBM are very similar (see Fig. 5B), for the velocity we use Eq. (\ref{lambda}) multiplied by $a v_{0}$,
\begin{equation}\label{vel_exp}
V(c,p) = a v_{0} \frac{2pc}{\sqrt{p^2(1-c)^2 + 4pc}},
\end{equation}
where $v_{0}$ is the transition rate. For the diffusion coefficient analytical results are not known, but for $c \ll 1$ the limiting result of $c=0$ can be used, yielding
\begin{equation}\label{dif_exp}
D=a^{2} v_{0}.
\end{equation}
Eqs. (\ref{vel_exp}) and (\ref{dif_exp}) with 2 unknowns, $p$ and $v_{0}$, can be easily solved, and we obtain 
\begin{equation}
p \simeq 0.283, \quad v_{0} \simeq 8 \times 10^{3} \mbox{ s}^{-1}.
\end{equation}
Our estimate of the burning probability $p \simeq 28.3 $$\%$ is larger than the result $p=10$$\%$ obtained in Monte Carlo simulations \cite{saffarian04}. However, the difference  can be attributed to different versions of BBM used in our theoretical analysis and in computer simulations \cite{saffarian04}, and  to the uncertainty in the determination of the step size $a$.

It is also interesting to analyze the efficiency of the motor protein driven by burning of bridges. Our theoretical analysis indicates that the motion of the particle in the system with burnt bridges can be viewed as a biased random walk. The dynamic properties of the random walker effectively can be written as
\begin{equation}\label{eqn_eff}
V=a(u_{eff}-w_{eff}), \quad D=\frac{1}{2}a^{2}(u_{eff}+w_{eff}),
\end{equation}
where $u_{eff}$ and $w_{eff}$ are effective transition rates of the biased random walk. Then the force exerted by the molecular motor can be found from \cite{AR07,KF}
\begin{equation}
F=\frac{k_{B}T}{a} \ln{\frac{u_{eff}}{w_{eff}}}.
\end{equation}
Substituting experimental values for $V$ and $D$ into Eqs. (\ref{eqn_eff}) to obtain the effective rates $u_{eff}$ and $w_{eff}$, the exerted force can be calculated, yielding $F \simeq 0.023$ pN, which is two orders of magnitude weaker than the forces produced by other motor proteins \cite{howard_book}. However, this force is exerted over the distance of $\Delta=300$ nm, which is 10-100 times larger than for conventional molecular motors, producing $W=F \times \Delta \simeq 6.9$ pN$\times$nm of work. The cleavage of one peptide bond, that fuels the motion of collagenase, liberates $E \simeq 20.9$ kJ/mole or 34.8 pN$\times$nm of energy \cite{saffarian04}. The calculated efficiency of the collagenase motor protein, $\nu=W/E \simeq 19.8$$\%$, is only slightly larger than the efficiency of 15$\%$ obtained from the experimental estimates  \cite{saffarian04}. Thus the collagenase is a weak molecular motor and its efficiency is not very high in comparison with other motor proteins.

\section{Summary and Conclusions}

A general theoretical analysis of particle dynamics in different one-dimensional  burnt-bridge models is presented. It is shown that when the probability of burning is equal to unity, the system can be mapped into the model of the single particle hopping along the periodic infinite lattice. This allows us to obtain all dynamic properties in exact and explicit form. For general case of $p< 1$ we developed  a method of analyzing the dynamics by considering a reduced kinetic scheme with respect to the last burnt bridge. It is found that distribution of weak links and the details of the burning mechanism strongly influence the particle dynamics in BBM. Theoretical calculations are fully supported by Monte Carlo computer simulations. Application of this theoretical method to experiments on collagenase motor proteins provides the estimates of the burning probability and the exerted force, leading to conclusion that this molecular motor is rather weak and not very efficient, in contrast to conventional motor proteins.

Although the presented theoretical approach  investigates explicitly the dynamic  properties of many BBMs and it is successfully applied  for analysis of experimental data, there are several problems that should be addressed in the future studies.  We have not been able to obtain expressions for the diffusion coefficient in the general case of $p < 1$. It is also not clear what is the origin of jump in the dispersion  in the limit of $c \rightarrow 0$. It will be interesting to investigate the motion of the motor protein on periodic lattices \cite{AR07} with  bridge burning, and it is reasonable to suggest that the approach developed in this paper can be generalized for this case too. In addition, a more realistic approach will be to allow the molecular motor to jump between parallel tracks \cite{saffarian04}. Future experiments on single motor proteins that utilize bridge  burning mechanisms will be important to test  the validity of the presented theoretical approach.

\section{Acknowledgments}

The authors would like to acknowledge the support from the Welch Foundation (grant C-1559), the U.S. National Science Foundation (grant  CHE-0237105) and Hammill Innovation Award.

\newpage

\noindent {\bf Figure Captions:} \\\\

\noindent Fig. 1. A schematic view of the motion of the particle in the system with burnt bridges. Thick solid lines are strong links, while thin solid lines are weak links, or ``bridges'' that can be burnt.  Dotted lines are already burnt bridges. The particle can jump with equal rates to the right or to the left unless the link is broken.

\vspace{5mm}

\noindent Fig. 2. Dynamic properties of continuous-time burnt-bridge model with periodic distribution of bridges for $p=1$. Lines are from analytic calculations, while symbols are obtained from Monte Carlo computer simulations - see text for details.

\vspace{5mm}

\noindent Fig. 3. Reduced kinetic schemes for continuous-time systems with periodic distribution of bridges: a) forward BBM (only transition rates not equal to unity are shown); b) forward-backward BBM (only transition rates not equal to 1/2 are shown). The origin corresponds to the right end of the last burnt bridge.

\vspace{5mm}

\noindent Fig. 4. A) Velocity as a function of concentration of bridges for different burning probabilities for the periodic forward continuous-time BBM. Solid lines are results of analytical calculations, and symbols are from Monte Carlo simulations. Curve a) corresponds to $p=0.9$; curve b) corresponds to $p=0.7$; curve c)  corresponds to $p=0.5$; and curve d) corresponds to $p=0.1$.\\
B) Velocity as a function of burning probability for different concentrations of weak links for the periodic forward continuous-time  BBM. Solid lines are results of analytical calculations, and symbols are from Monte Carlo simulations. Curve a) corresponds to $c=1$; curve b) corresponds to $c=1/2$; curve c)  corresponds to $c=1/3$; and curve d) corresponds to $c=1/10$.

\vspace{5mm}

\noindent Fig. 5. A) Velocity as a function of concentration for different distributions of bridges for the periodic forward continuous-time  BBM. Solid lines are results of analytical calculations, and symbols are from Monte Carlo simulations. Curve a) corresponds to random distribution with $p=1$; curve b) corresponds to periodic distribution  with $p=1$; curve c)   corresponds to random distribution with $p=0.5$; and curve d) corresponds to periodic distribution with $p=0.5$. \\
B) Velocity as a function of concentration for different burning realizations for the periodic  continuous-time periodic BBM. Solid lines are results of analytical calculations. Curve a) corresponds to forward-backward BBM with $p=0.8$; curve b) corresponds to forward BBM with $p=0.8$; curve c) corresponds to forward-backward BBM with $p=0.2$; and  curve d) 
corresponds to forward BBM with $p=0.2$.

\vspace{5mm}

\noindent Fig. 6. Dispersion in the periodic forward continuous-time periodic BBM from Monte Carlo simulations: A) as a function of burning probability; and B) as a function of concentration of bridges.

\newpage

\noindent \\\\\\

\begin{figure}[ht]
\unitlength 1in
\resizebox{3.375in}{0.65in}{\includegraphics{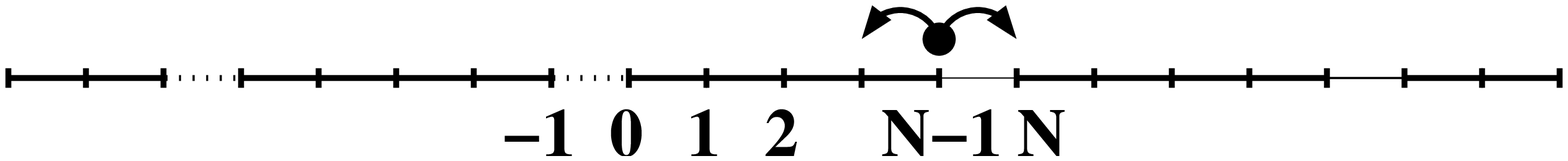}}
\vskip 0.3in
\begin{Large} Fig.1   \end{Large}
\end{figure}

\newpage

\noindent \\\\\\

\begin{figure}[ht]
\begin{center}
\unitlength 1in
\resizebox{3.375in}{3.0in}{\includegraphics{Fig2.eps}}
\vskip 0.3in
 \begin{Large} Fig.2   \end{Large}
\end{center}
\end{figure}

\newpage

\noindent \\\\\\

\begin{figure}[ht]
\begin{center}
\unitlength 1in
\resizebox{3.375in}{2.70in}{\includegraphics{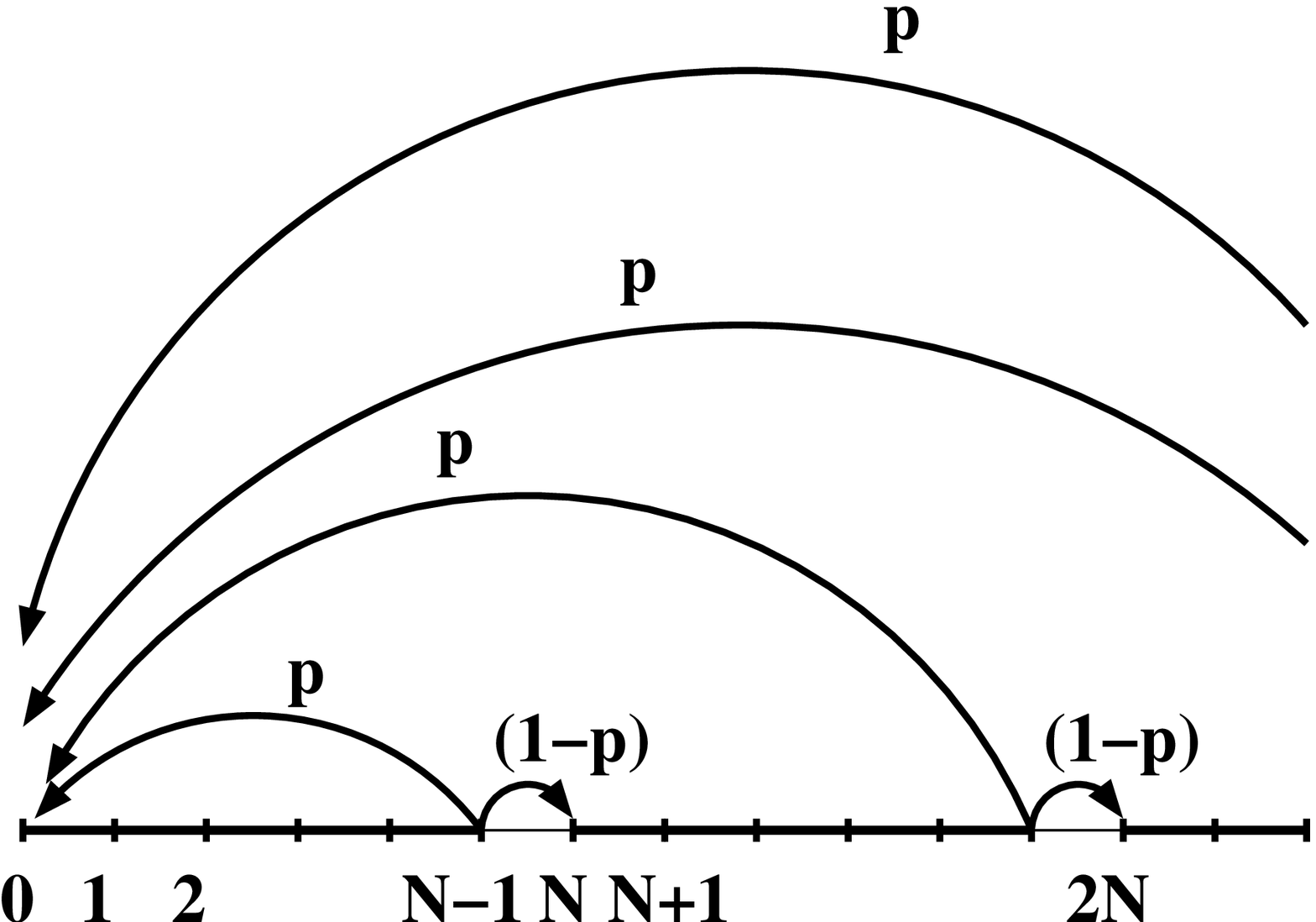}}
\vskip 0.3in
 \begin{Large} Fig.3a   \end{Large}
\end{center}
\end{figure}

\newpage

\noindent \\\\\\

\begin{figure}[ht]
\begin{center}
\unitlength 1in
\resizebox{3.375in}{2.70in}{\includegraphics{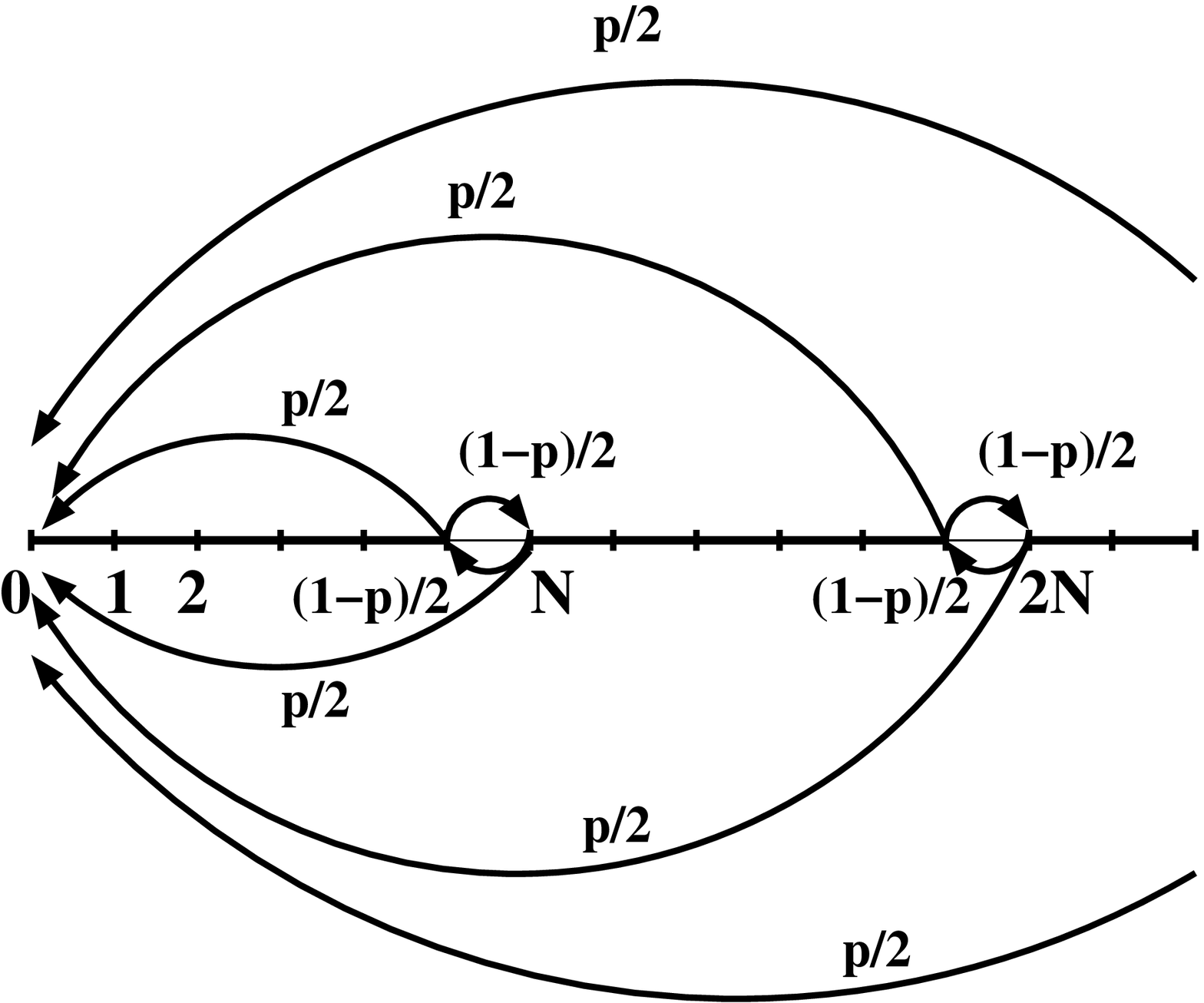}}
\vskip 0.3in
 \begin{Large} Fig.3b  \end{Large}
\end{center}
\end{figure}

\newpage

\noindent \\\\\\

\begin{figure}[ht]
\begin{center}
\unitlength 1in
\resizebox{3.375in}{2.70in}{\includegraphics{Fig4a.eps}}
\vskip 0.3in
 \begin{Large} Fig.4A  \end{Large}
\end{center}
\end{figure}

\newpage

\noindent \\\\\\

\begin{figure}[ht]
\begin{center}
\unitlength 1in
\resizebox{3.375in}{2.70in}{\includegraphics{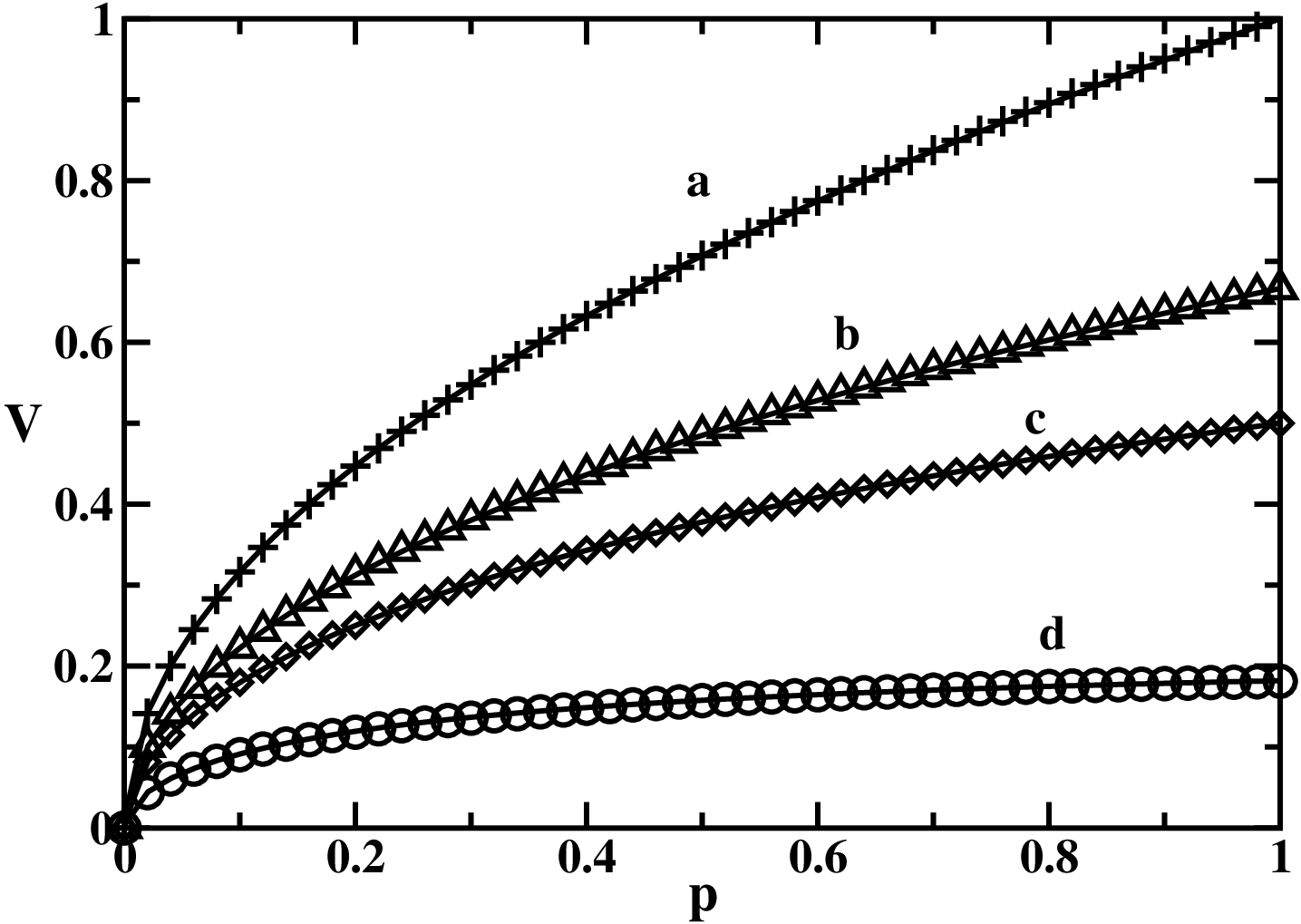}}
\vskip 0.3in
 \begin{Large} Fig.4B  \end{Large}
\end{center}
\end{figure}

\newpage

\noindent \\\\\\

\begin{figure}[ht]
\begin{center}
\unitlength 1in
\resizebox{3.375in}{2.70in}{\includegraphics{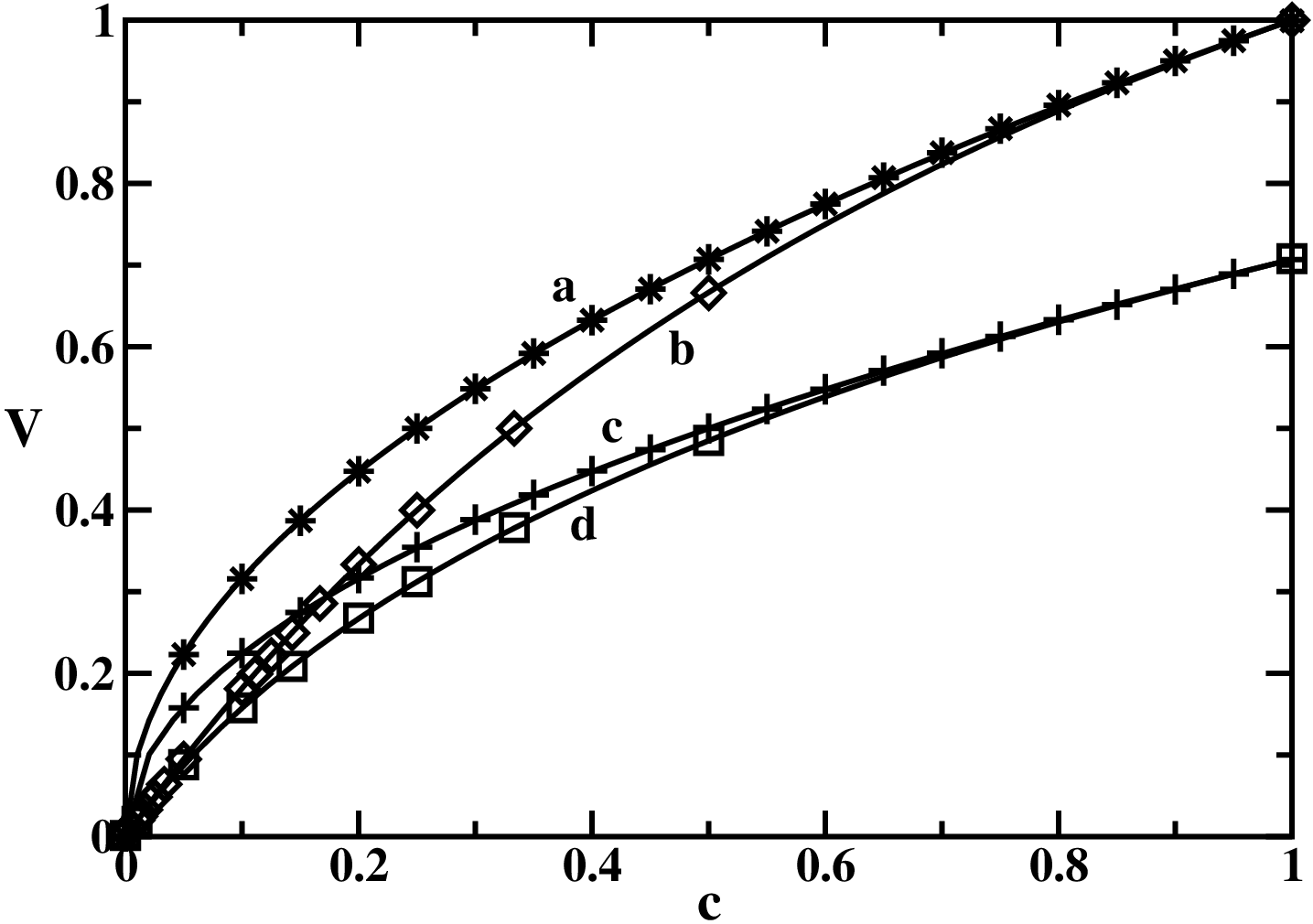}}
\vskip 0.3in
 \begin{Large} Fig.5A  \end{Large}
\end{center}
\end{figure}

\newpage

\noindent \\\\\\

\begin{figure}[ht]
\begin{center}
\unitlength 1in
\resizebox{3.375in}{2.70in}{\includegraphics{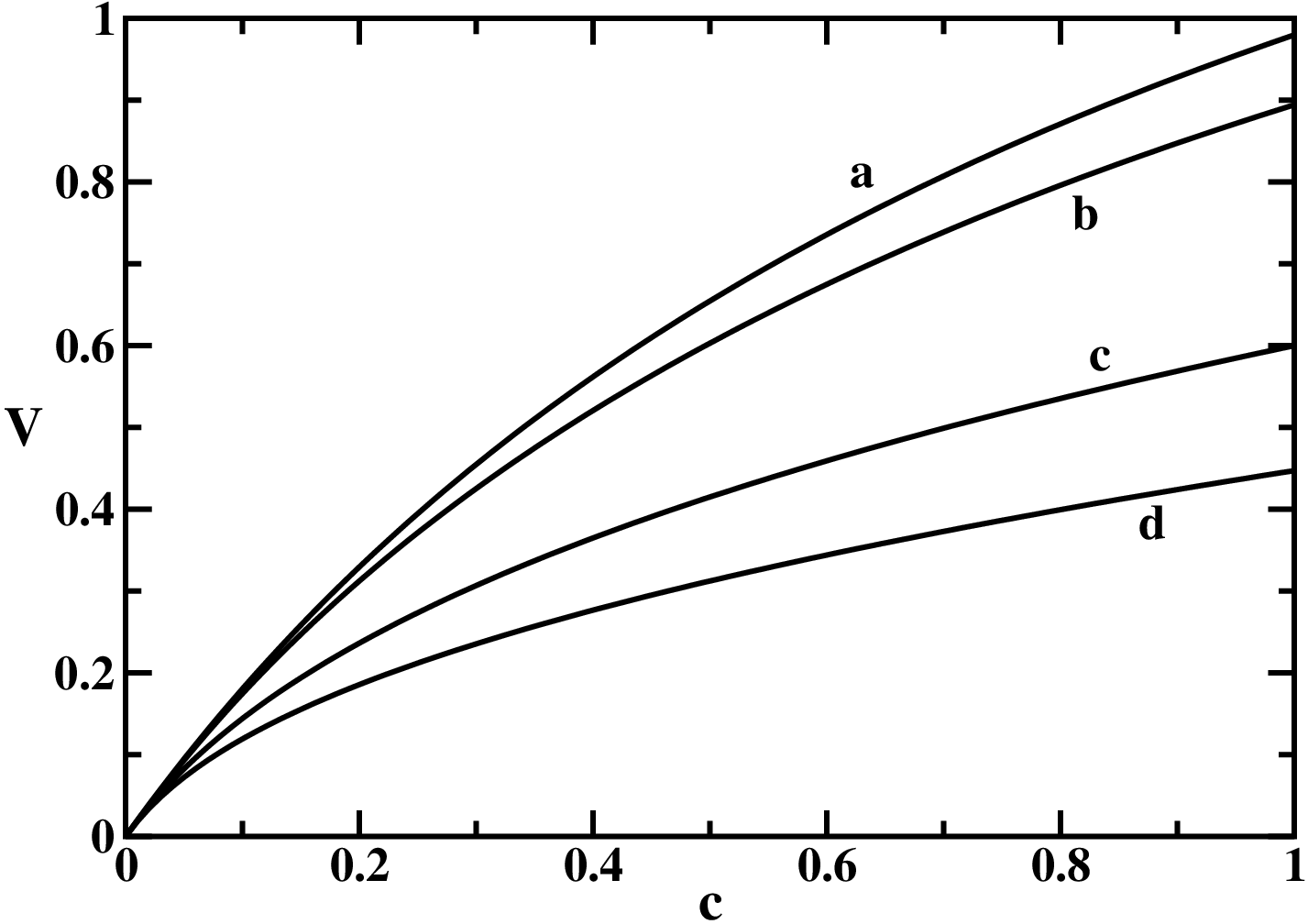}}
\vskip 0.3in
 \begin{Large} Fig.5B  \end{Large}
\end{center}
\end{figure}

\newpage

\noindent \\\\\\

\begin{figure}[ht]
\begin{center}
\unitlength 1in
\resizebox{3.375in}{2.70in}{\includegraphics{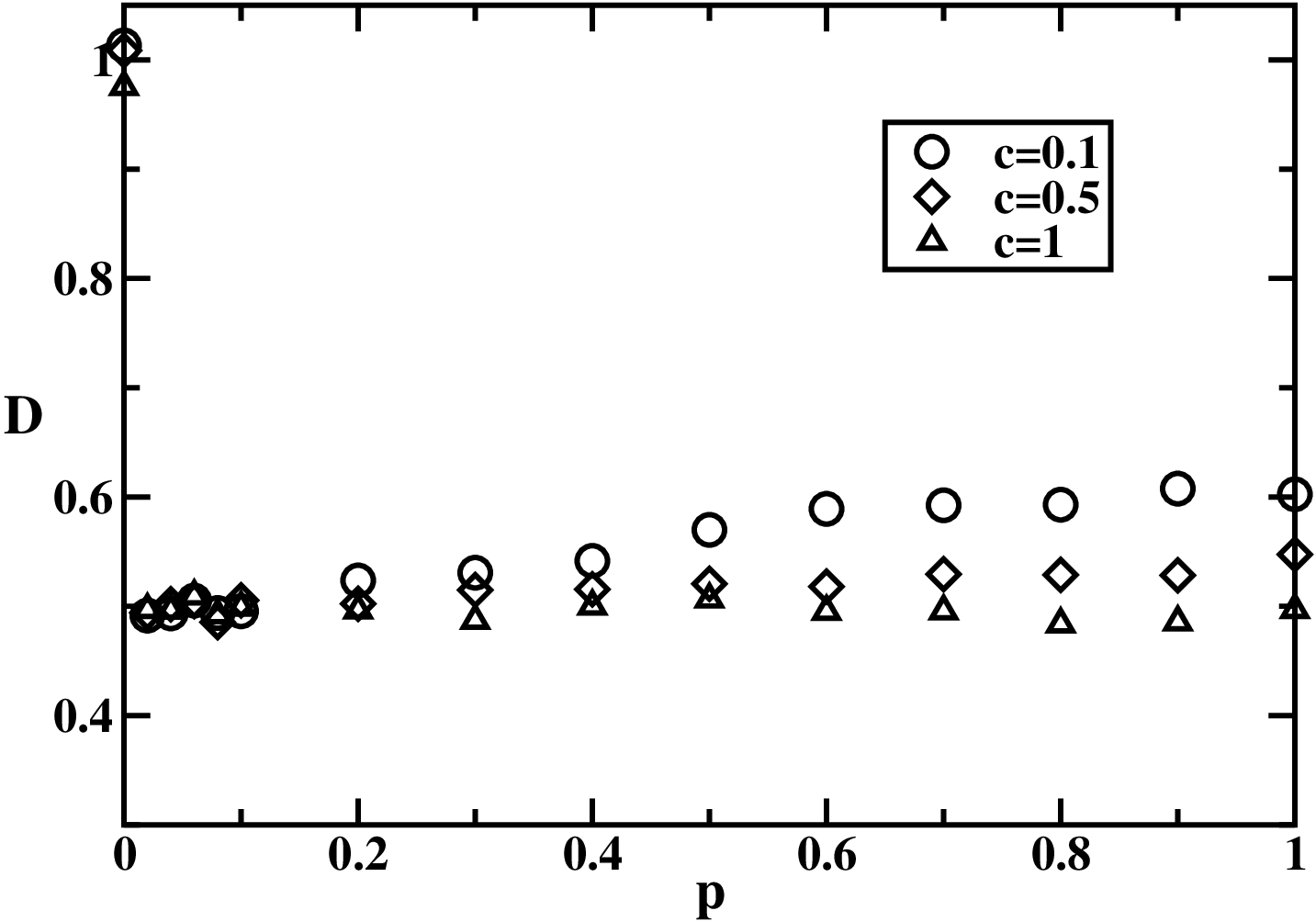}}
\vskip 0.3in
 \begin{Large} Fig.6A  \end{Large}
\end{center}
\end{figure}

\newpage

\noindent \\\\\\

\begin{figure}[ht]
\begin{center}
\unitlength 1in
\resizebox{3.375in}{2.70in}{\includegraphics{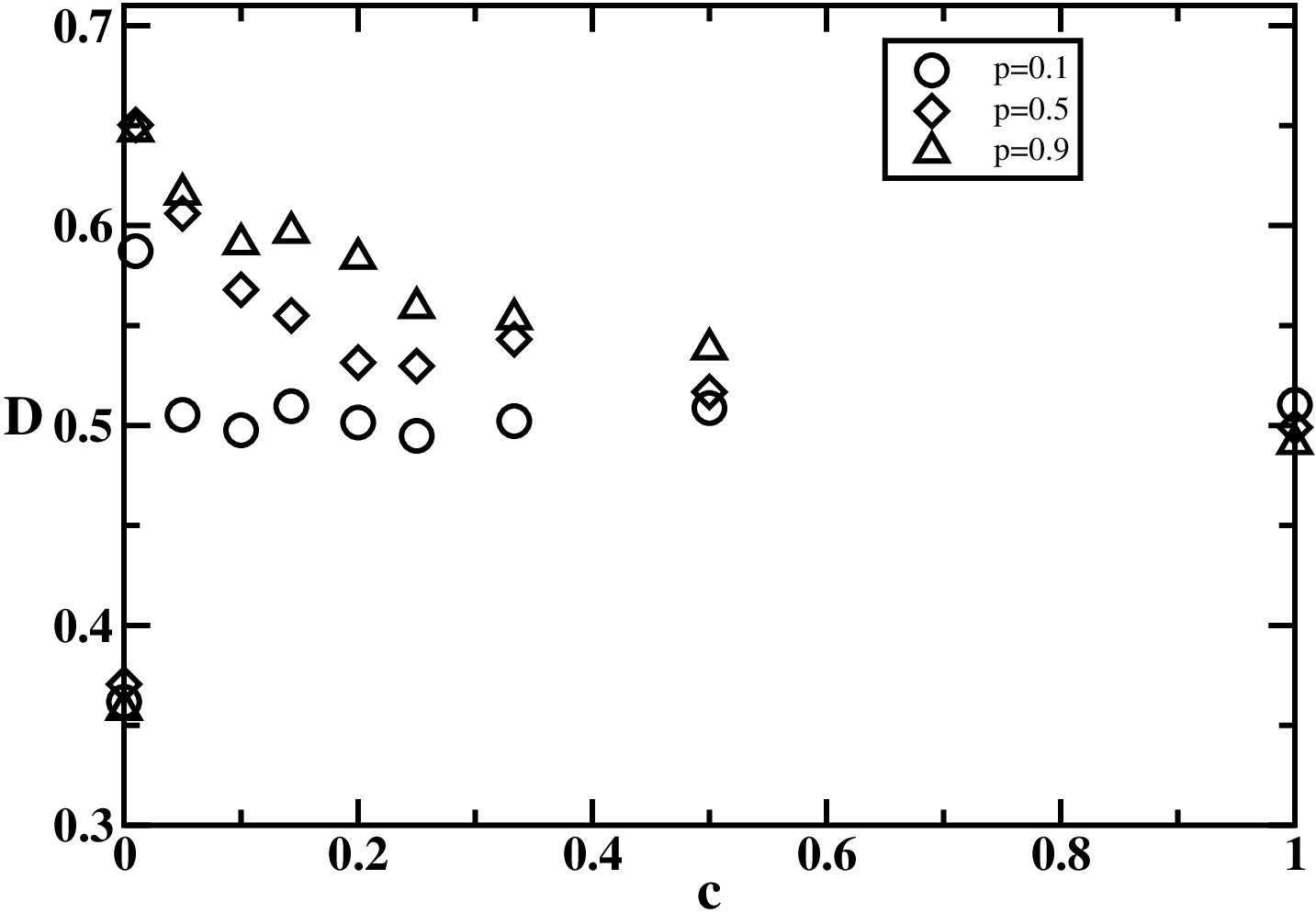}}
\vskip 0.3in
 \begin{Large} Fig.6B  \end{Large}
\end{center}
\end{figure}


\begin{thebibliography}{99}

\bibitem{lodish_book} H. Lodish et. al., {\it Molecular Cell Biology}, (W.H. Freeman and Company, New York, 2000).

\bibitem{howard_book} J. Howard, {\it Mechanics of Motor Proteins and Cytoskeleton}, (Sinauer Associates, Sunderland Massachusetts, 2001).

\bibitem{bray_book} D. Bray, {\it Cell Movements. From Molecules to Motility}, (Garland Publishing, New York, 2001). 

\bibitem{saffarian04} S. Saffarian, I.E. Coller, B.L. Marmer, E.L. Elson, and G. Goldberg, Science {\bf 306}, 108 (2004). 

\bibitem{saffarian06} S. Saffarian, H. Qian, I.E. Coller, E.L. Elson, and G. Goldberg, Phys. Rev. E {\bf 73}, 041909 (2006). 

\bibitem{mai01} J. Mai, I.M. Sokolov, and A. Blumen, Phys. Rev. E {\bf 64}, 011102 (2001).

\bibitem{antal05} T. Antal and P.L. Krapivsky, Phys. Rev. E {\bf 72}, 046104 (2005).

\bibitem{kolomeisky00} A.B. Kolomeisky and M.E. Fisher, J. Chem. Phys. {\bf 113}, 10867 (2000).

\bibitem{derrida83} B. Derrida, J. Stat. Phys. {\bf 31}, 433 (1983).

\bibitem{AR07} A.B. Kolomeisky and M.E. Fisher, to appear in Ann. Rev. Phys. Chem, (2007).

\bibitem{mccarthy93} J. F. McCarthy, J. Phys. A: Math. Gen. {\bf 26}, 2495 (1993).

\bibitem{KF} M.E.Fisher and A.B. Kolomeisky, Proc. Natl. Acad. Sci. USA  {\bf 96}, 6597 (1999).

\end{thebibliography}
\end{document}